\documentclass[prl,twocolumn,showpacs,preprintnumbers,amsmath,amssymb]{revtex4}
\usepackage{amsfonts,amssymb}
\usepackage{graphicx}

\begin{document}
\draft
\title{MAGNONIC CRYSTAL THEORY OF THE SPIN-WAVE FREQUENCY GAP IN LOW-DOPED $La_{1-x}Ca_{x}MnO_{3}$ MANGANITES}
\author{M. Krawczyk, H. Puszkarski}
\address{Surface Physics Division, Faculty of Physics, Adam
Mickiewicz University,  ul. Umultowska 85, Pozna\'{n}, 61-614 Poland.}

\date{\today}

\begin{abstract}
A theory of three-dimensional (3D) hypothetical magnonic crystal (conceived as the magnetic counterpart of the well-known photonic crystal) is
developed and applied to explain the existence of a spin-wave frequency gap recently revealed in low-doped manganites $La_{1-x}Ca_{x}MnO_{3}$ by
neutron scattering. A successful confrontation with the experimental results allows us to formulate a working hypothesis that certain manganites
could be regarded as 3D magnonic crystals existing in nature.
\end{abstract}
 \maketitle

\section{Introduction}
Though the first study of electromagnetic wave propagation in periodic structures, written by lord Rayleigh, was published already in 1887, not
until very recently have photonic composites suddenly raised an extremely keen interest. The research in this field was triggered by the studies
by Yablonovitch and John \cite{[1],[2]}, published in 1987 and anticipating the existence of complete energy gaps in electromagnetic wave
spectra in three-dimensional periodic composites, henceforth referred to as {\em photonic crystals}. These can be used for fabricating new
optoelectronic devices with photons acting as transport medium. The so-called {\em left-handed materials}, an example of periodic structures
characterized by negative effective refractivity \cite{[7]}, demonstrate how much the properties of this kind of structure can differ from those
of homogeneous materials. Another type of periodic composites are structures composed of materials with different elastic properties; showing an
energy gap in their elastic wave spectrum, such composites are referred to as {\em phononic crystals} \cite{[8]}. Recently, attention has been
focused on the search of photonic and phononic crystals in which both the position and the width of the energy gap could be controlled by
external factors, such as applied voltage or {\em magnetic field}. Attempts are made to create photonic crystals in which one of the component
materials would be a magnetic \cite{[12],[15]}.

A magnetic periodic composite, conceived as the magnetic counterpart of a photonic crystal, consists of at least two magnetic materials, with
magnon acting as information carrier, therefore such periodic magnetic composites can be referred to as {\em magnonic crystals} (MC). Studies of
2D magnonic crystals have already been reported \cite{[16],[17],[19],[20],[21]} with scattering centers in the form of "infinitely" long
cylinders disposed in square lattice nodes (cylinder and matrix materials being two different ferromagnetics) and spin wave spectra showing the
anticipated gaps. Further research was focused on magnetic multilayer systems, which can be regarded as 1D magnonic crystals \cite{[24],[30]}.
In this paper, we present numerically calculated - and, to our best knowledge, not yet reported in the literature - band structures of {\em
three-dimensional} magnonic crystals. Due to the complexity of the problem, only the simplest model of 3D magnonic crystal is considered here,
represented by a system of ferromagnetic spheres (which act as scattering centers) disposed in the nodes of a cubic {\em bcc} crystal lattice
and embedded in a different ferromagnetic material (matrix). Both the exchange and dipolar interactions are taken into account in our
calculations, which are based on the plane wave method and use the linear approximation. As a conclusion, we propose a new magnonic
interpretation of experimental results obtained through neutron scattering on spin waves in doped manganites
\cite{hennion98,biotteau01,kober04}.

\section{Theory of 3D magnonic band structure}
\begin{figure}[h]
\begin{center}
\includegraphics[width=60mm]{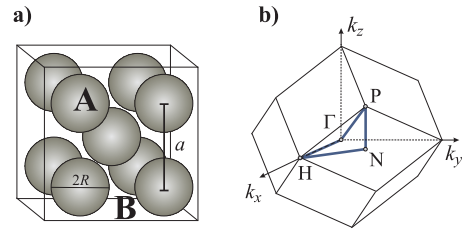}
\vspace{2mm} \caption{a) The 3D periodic structure studied in this paper; the structure consists of ferromagnetic material {\bf A} spheres
embedded in a material {\bf B} matrix (materials {\bf A} and {\bf B} having different magnetic properties) and disposed in the nodes of a {\em
bcc} lattice. (b) The Brillouin zone corresponding to the considered structure, and the path ($P \Gamma HNP$, highlighted) along which the
magnonic band spectra are calculated. \label{fig1}}
\end{center} \end{figure}

Let's consider an ideal periodic structure consisting of spheres of ferromagnetic {\bf A} embedded in a matrix of ferromagnetic {\bf B}. The
spheres are assumed to form a 3D periodical lattice of {\em bcc} (Fig. 1a) type. A static magnetic field, $H_{0}$, is applied to the composite
along the $z$ axis, and assumed to be strong enough to saturate the magnetization of both materials. The lattice constant is denoted by $a$; the
filling fraction, $f =\frac{8}{3} \pi R^{3} a^{-3}$, is defined as the volume proportion of material {\bf A} in a unit cell. Ferromagnetics {\bf
A} and {\bf B} are characterized by two material parameters: the spontaneous magnetization ($M_{S,A}$ and $M_{S,B}$), and the exchange constant
($A_{A}$ and $A_{B}$); both these parameters depend on the position vector $\vec{r}=(x,y,z)$:
\begin{eqnarray}
M_{S}(\vec{r})&=&M_{S,B}+(M_{S,A}-M_{S,B})S(\vec{r}),
\nonumber \\
A(\vec{r})&=&A_{B}+(A_{A}-A_{B})S(\vec{r}),\label{eq2}
\end{eqnarray}
the value of function $S(\vec{r})$ being 1 inside a sphere, and 0 beyond.

In the classical approximation, spin waves are described by the Landau-Lifshitz (LL) equation, taking the following form in the case of magnetic
composites (with damping neglected):
\begin{eqnarray}
\frac{\partial \vec{M}(\vec{r},t)}{\partial t} = \gamma \mu_{0}
\vec{M}(\vec{r},t) \times \vec{H}_{eff}(\vec{r},t), \label{eq3}
\end{eqnarray}
where magnetization $\vec{M}(\vec{r},t)$ is a function of position vector $\vec{r}$ and time $t$; $\vec{H}_{eff}(\vec{r},t)$ stands for the
effective magnetic field  \cite{landau35,[16],[17],[19]} acting on magnetization $\vec{M}(\vec{r},t)$:
\begin{eqnarray}
\vec{H}_{eff}(\vec{r},t)=H_{0}\hat{z}
+\vec{h}(\vec{r},t)+\frac{2}{\mu_{0}} \left(\nabla \cdot
\frac{A}{M^{2}_{S}}\nabla \right) \vec{M}(\vec{r},t); \label{eq4}
\end{eqnarray}
$\hat{z}$ is the unit vector along the $z$ axis; $\vec{h}(\vec{r}, t)$ is the dynamic magnetic field resulting from dipolar interactions; the
third component represents the exchange field. The magnetization vector can be represented as the sum of its static and dynamic components; the
former, $M_{S}\hat{z}$, is parallel to the applied field; the latter, $\vec{m}(\vec{r},t)=\vec{m}(\vec{r})\exp(-i \omega t)$,  $\omega$ denoting
the precession circular frequency, lies in the plane ($x,y$):
\begin{equation}
\vec{M}(\vec{r},t)=M_{S}\hat{z}+\vec{m}(\vec{r},t). \label{eq5}
\end{equation}
The dynamic dipolar field, $\vec{h}$, must satisfy the magnetostatic Maxwell equations:
\begin{eqnarray}
\nabla \times \vec{h}(\vec{r})=0, \nonumber \\
\nabla \cdot \left(\vec{h}(\vec{r})+ \vec{m}(\vec{r})\right)=0.
\label{eq6}
\end{eqnarray}
In magnonic crystals, the position-dependent coefficients in (\ref{eq4}), {\em i.e.} $M_{S}$ and $A$, are periodic functions of the position
vector, which allows us to use in the procedure of solving the LL equation (\ref{eq3}) the plane wave method, described in detail in our earlier
papers \cite{[17],[19]} (dealing with {\em two-dimensional} magnonic crystals). Following this scheme, we proceed to Fourier-expanding all the
periodic functions of the position vector, {\em i.e.} the spontaneous magnetization, $M_{S}$, and parameter $Q$ defined as follows:
\begin{eqnarray}
Q=\frac{2A}{\mu_{0}M^{2}_{S}H_{0}}. \label{eq7}
\end{eqnarray}
The dynamic components of the magnetization can be expressed as the product of the periodic envelope function and the Bloch factor, $exp(i\vec{k}\vec{r}$) ($\vec{k}$ denoting a 3D wave vector); the envelope function can be transformed into the reciprocal space as well. Including all the expansions into (\ref{eq3}) and (\ref{eq6}) leads to the following infinite system of linear equations for Fourier coefficients of the dynamic magnetization components, $\vec{m}_{x \vec{k}}(\vec{G})$  and $\vec{m}_{y \vec{k}}(\vec{G})$:
\begin{widetext}
\begin{eqnarray}
  i \Omega  m_{x \vec{k}}(\vec{G})= m_{y \vec{k}}(\vec{G}) +\sum_{\vec{G}'} \frac{(k_{y} + G'_{y})(k_{x}+G'_{x}) m_{x\vec{k}}(\vec{G}') +
(k_{y} + G'_{y})^{2}m_{y \vec{k}}(\vec{G}')}{H_{0}| \vec{k}+
 \vec{G}' |^{2}} M_{S}(\vec{G}- \vec{G}')  \nonumber \\
  +\sum_{\vec{G}'} \sum_{\vec{G}''}[ (\vec{k} + \vec{G}') \cdot (\vec{k}+\vec{G}'')
- (\vec{G}-\vec{G}'') \cdot (\vec{G} - \vec{G}' ) ]
  M_{S}(\vec{G}-\vec{G}'') Q(\vec{G}'' - \vec{G}') m_{y \vec{k}} (\vec{G}'),\nonumber
\\
 i \Omega m_{y \vec{k}}(\vec{G})= -m_{x \vec{k}} (\vec{G})-\sum_{\vec{G}'}
\frac{(k_{y} + G'_{y})(k_{x}+G'_{x}) m_{y \vec{k}}(\vec{G}')+(k_{x} + G'_{x})^{2}m_{x \vec{k}}(\vec{G}')}{H_{0}| \vec{k} + \vec{G}' |^{2}}
 M_{S}(\vec{G} -\vec{G}') \nonumber \\
- \sum_{\vec{G}'} \sum_{\vec{G}''} [ (\vec{k} + \vec{G}') \cdot (\vec{k}+\vec{G}'') - (\vec{G}-\vec{G}'') \cdot (\vec{G} - \vec{G}' ) ]
M_{S}(\vec{G}-\vec{G}'') Q(\vec{G}'' - \vec{G}') m_{x \vec{k}} (\vec{G}'); \nonumber \\ \label{eq8}
\end{eqnarray}
\end{widetext}
$k_{x},k_{y}$ and $G_{x},G_{y}$ denoting the Cartesian components of the wave vector $\vec{k}$ and a reciprocal lattice vector $\vec{G}$,
respectively; a new quantity introduced in (\ref{eq8}) is $\Omega$, henceforth referred to as {\em reduced frequency}:
\begin{eqnarray}
\Omega=\frac{\omega}{|\gamma| \mu_{0} H_{0}}. \label{eq9}
\end{eqnarray}
The Fourier coefficients of spontaneous magnetization $M_{S}$ and
parameter $Q$ are calculated from the inverse Fourier
transformation; in the case of spheres, the resulting formulae for $M_{S}$ read
 as follows:
\begin{eqnarray*}
M_{S}(\vec{G}) = \left\{ \begin{array}{l}
     M_{S,A}f + M_{S,B}(1-f),\;
     \mbox{for  $\vec{G}=0$} \\
f  (M_{S,A}-M_{S,B})
{\displaystyle \frac{3\left[\sin(GR)-(GR)\cos(GR)\right]}{(GR)^{3}}},\\
\mbox{ \hspace{5cm}for $\;\vec{G} \neq 0$}%
    \end{array}
    \right.%
     \label{aa1}
\end{eqnarray*}
and similarly for $Q$; $R$ is the sphere radius (Fig.
\ref{fig1}a).

Obviously, the numerical calculations performed on the basis of
(\ref{eq8}) involve a finite number of reciprocal lattice vectors
$\vec{G}$ in the Fourier expansions; however, we have made sure
that the number used is large enough to guarantee good convergence of the numerical results. As indicated by an analysis performed, a satisfactory convergence is obtained already with 343 reciprocal lattice vectors used.

\section{Numerical results and confrontation with the experiment in manganites}
\begin{figure}[h]
\begin{center}
\includegraphics[width=80mm]{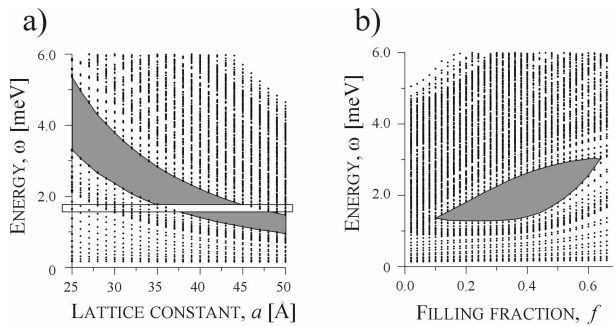}
\caption{ Magnonic bands computed on the basis of (\ref{eq8}) for a hypothetic \textit{periodic} arrangement of ferromagnetic droplets embedded
in different ferromagnetic medium; the sphere-shaped droplets are assumed to form a regular crystal lattice of {\em bcc} type. (a) Magnonic
branches plotted versus the lattice constant, $a$, of the droplet {\em bcc} structure; the filling fraction is fixed at $f = 0.33$. A frequency
gap (the shaded region) is found to exist, and move down with increasing $a$. The superimposed empty rectangle represents the gap revealed
experimentally in doped manganites \cite{hennion98}. (b) The magnonic energy branches (in the same \textit{crystal}) plotted \textit{versus} the
filling fraction, with the lattice constant fixed at $a=42\AA$.} \label{fig2}
\end{center} \end{figure}

An energy gap in spin wave spectra has recently been found  experimentally in doped manganites $La_{1-x}Ca_{x}MnO_{3}$ and
$La_{1-x}Sr_{x}MnO_{3}$ at low ($Ca$ or $Sr$) ion concentrations ($2\% <x \leq 10\%$) \cite{hennion98,biotteau01}. The appearance of this gap
was suggested to be due to doping, which results in a two-phase magnetic structure, with ferromagnetic (F) droplets (associated with the
$Mn^{4+}$ ions) embedded in a canted antiferromagnetic (CAF) medium (showing a non-zero magnetic moment along the direction of the droplet
magnetic moments). Hence, the two {\em separate} energy branches found in the neutron scattering experiments are interpreted as corresponding to
two {\em independent} spin-wave excitations, each propagating in one of the two phases. In this paper, we propose an alternative {\em magnonic}
explanation of the spin-wave spectrum energy gap found in these experiments. We anticipate that the dispersion branches obtained in the
experiment constitute the bottom part of the {\em collective} spin-wave spectrum, involving spin waves propagating {\em concurrently through
both phases} of a magnetic composite, resulting from a {\em regular} arrangement of ferromagnetic droplets (material {\bf A}) in a matrix with
different magnetic properties (material {\bf B}). Then, the existence of a spin-wave spectrum gap becomes a natural consequence of the structure
periodicity, as in the case of photonic or phononic crystals. Below we present results of our calculations (based on the method discussed in the
preceding paragraph) of the spin-wave spectrum in a {\em droplet magnonic crystal}, the necessary material parameter values being as in the
experimental studies \cite{hennion98,biotteau01}. The results are compared to the spectrum obtained experimentally in $La_{0.9}Ca_{0.1}MnO_{3}$.

Let's assume that the spherical ferromagnetic droplets form a {\em bcc} lattice (Fig. \ref{fig1}a), which we believe to be a good approximation
of the actual droplet structure. The spin-wave branches will be computed along the following path in the irreducible part of the first Brillouin
zone: $P=\pi/a(1,1,1) \rightarrow \Gamma= \pi/a(0,0,0) \rightarrow H=\pi/a(2,0,0) \rightarrow N=\pi/a(1,1,0) \rightarrow P=\pi/a(1,1,1)$ (see
Fig. \ref{fig1}b). The necessary material parameter values, {\em i.e.} those of the spontaneous magnetization and the exchange constant, in both
magnetic phases of our model magnonic crystal are as in the experimental studies referred to above. The paper by Hennion {\em et al.}
\cite{hennion98} provides information on the difference, $\Delta m$, between the values of magnetization along vector $\vec{c}$ in the droplets
and in the matrix: $\Delta m = 0.7 \pm 0.2 \mu_{B}$. Assuming that the magnetic moment in droplets comes mainly from $Mn^{4+}$ ions with spin $S
= 3/2$, we get droplet spontaneous magnetization value $M_{S,A}=0.46\; 10^{6}Am^{-1}$; consequently, the matrix magnetization ($M_{S,A}$ less
0.7$\mu_{B}\equiv 0.22 Am^{-1}$) is $M_{S,B}=0.24\; 10^{6}Am^{-1}$. The following lattice constant values in $La_{0.9}Ca_{0.1}MnO_{3}$ are
assumed: $a = 5.465\AA$, $b = 5.621\AA$ , and $c = 7.725\AA$ \cite{kober04}. The exchange constant values in droplets and in the matrix, $A_{A}$
and $A_{B}$, respectively, are determined on the basis of the exchange integral, $J_{1}(F)=1meV$, and the stiffness constant, $D \approx
15meV\AA^{2}$, as estimated in the study by Biotteau {\em et al.} \cite{biotteau01}; the resulting droplet exchange constant value is
$A_{A}=S^{2}J_{1}(F)/a_{c}\;[meV \AA^{-1}]=0.09 \cdot 10^{-11} \; Jm^{-1}$, $a_{c}$ being the lattice constant recalculated for the manganite
pseudocubic elementary cell. The matrix exchange constant value, $A_{B}$, is determined by fitting the weighted mean value of the droplet and
matrix stiffness constants to the stiffness constant value estimated in \cite{biotteau01}. We assume $A_{B}=0.012 \;Jm^{-1}$, which, at filling
fraction $f = 0.2$, corresponds to mean value $D=18.31 \; meVA^{2}$, close to the value estimated by fitting to the experimental results at
temperature $15^{0}$K. The assumed average value of the isotropic internal field, $\mu_{0}H=0.1 T$, lies between the values found by Savosta
{\em et al.} \cite{savosta01} (0.064T) and Yates {\em et al.} \cite{yates03} (0.35T).

\begin{figure}[h]
\begin{center}
\includegraphics[width=80mm]{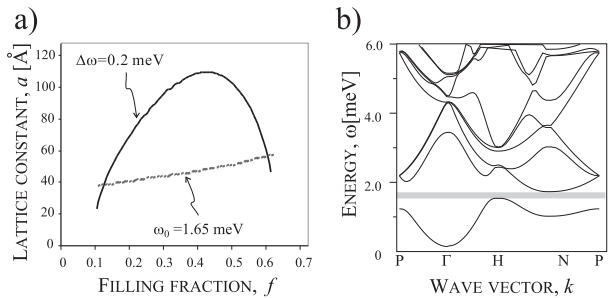}
\caption{(a) The $(a, f)$ plane with two sets of lattice constant $a$ and filling fraction $f$ values in a magnonic crystal: one, represented by
the {\em solid} line, corresponds to an energy gap whose width is exactly $\Delta \omega = 0.2meV$; the other, represented by the {\em dotted}
line, corresponds to a gap centered around $\omega_{0} = 1.65 meV$. The cross-section points indicate the $a$ and $f$ values at which the
computed gap fits the experimental one in terms of both width ($\Delta \omega$) and position ($\omega_{0}$). (b) Magnonic bands calculated on
the basis of our model across the Brillouin zone, the assumed lattice constant and filling fraction values being $a = 38.5\AA$, and $f = 0.13$,
respectively. The gap which is found to exist has position and width exactly fitting the experimental data reported in \cite{biotteau01}.}
\label{fig3}
\end{center} \end{figure}

Figure \ref{fig2}a shows the computed magnonic band structure plotted against the droplet lattice constant. The gap center ($\omega_{0}$) is
found to descend, and the gap itself (the dark gray region) to narrow down as the lattice constant increases. The superimposed rectangle
represents the 'experimental' gap reported in \cite{biotteau01}). The {\em gap center} found experimentally corresponds to our results at $a =
42 \AA$; however, the corresponding gap width resulting from our computations is much above the experimental result, which implies the necessity
of fitting the filling fraction as well. Fig. \ref{fig2}b shows the magnonic band structure plotted against the filling fraction $f$ (the
lattice constant being fixed at $a = 42\AA$). The gap is found to open at $f = 0.10$ to reach its maximum width at $f \approx 0.42$ and vanish
at $f = 0.64$, which means that one gap width value corresponds to two different filling fraction values. Note also that the gap center moves up
as the filling fraction increases.
 Hence, the $a$ and $f$ values corresponding to an energy gap that fits the experimental results will be
determined in two steps (see Fig. \ref{fig3}a): first, we shall find the set of $a$ and $f$ values corresponding to the experimental gap width,
$\Delta \omega= 0.2meV$ (the solid line in Fig \ref{fig3}a); next, another line (the dotted one in the Figure) is plotted, corresponding to the
gap centered at the experimental value, $\omega_{0} = 1.65 meV$. The cross section of these two lines indicates the $a$ and $f$ values at which
both $\omega_{0}$ and $\Delta \omega$ are fitted. As the lines cross at two points: ($a = 38.5\AA$, $f = 0.13$) and ($a = 56.5\AA$, $f = 0.61$),
two solutions are obtained, one corresponding to the droplet radius $R = 9.6\AA$, the other to $R = 23.6\AA$. The former solution proves close
beyond expectation to the experimental result reported in \cite{biotteau01} (Fig. 9), where $La_{0.9}Ca_{0.1}MnO_{3}$ droplet radius is
estimated at $R = 9.37\AA$. The corresponding mean droplet spacing estimated in \cite{hennion98} is $d_{m} \approx 36\AA$, also quite close to
our result, $a = 38.5\AA$. Therefore, the complete set of structural and material parameters characterizing our droplet magnonic model of
$La_{0.9}Ca_{0.1}MnO_{3}$ is: $M_{S,A}=0.46\; 10^{6}Am^{-1}$, $M_{S,B}=0.24\; 10^{6}Am^{-1}$, $A_{A}=0.09 \; 10^{-11} Jm^{-1}$, $A_{B}=0.012\;
10^{-11}Jm^{-1}$, $\mu_{0}H=0.1 T$, $a=38.5\AA$, $f=0.13$. The corresponding full spin-wave spectrum is shown in Fig. \ref{fig3}b. An energy
gap, existing indeed between the first band and the second one, corresponds very well, in terms of both position and width, to that found
experimentally \cite{biotteau01}.

This result allows to propose a working hypothesis that low-doped manganites could be regarded as magnonic crystals existing in nature.
Obviously, further research is required for verification of this hypothesis; therefore, in another paper, we are going to investigate the effect
of non-spherical droplet shape on the magnonic spectrum, and to consider a real canted antiferromagnetic matrix, rather than the effective
ferromagnetic one considered here. However, already the results obtained here on the basis of the approximate model are extremely promising.

\subsection*{Acknowledgements}

The present work was supported by the Polish State Committee for
Scientific Research through projects KBN - 2P03B 120 23 and
PBZ-KBN-044/P03-2001.


\end{document}